\documentclass[3p]{elsarticle}

\usepackage{booktabs} 
\usepackage{array}

\newcommand{\srcstyle}[1]{\ttfamily\textbf{#1}\rmfamily}

\usepackage{comment}

\usepackage{xcolor}
\definecolor{black}{RGB}{0,0,0}

\newlength{\MaxSizeOfLineNumbers}%
\usepackage{listings}
\usepackage[utf8]{inputenc}
\usepackage{graphicx}
\usepackage{url}
\usepackage{hyperref}
\usepackage{pgf-umlsd}

\newcommand{\titulo}{Transparent Replication Using Metaprogramming in Cyan}

\hypersetup{
     pdftitle={\titulo},
     pdfauthor={Fellipe A. Ugliara, Gustavo M. D. Vieira, José de
     O. Guimarães},
     pdfdisplaydoctitle=true,
     hidelinks
}

\journal{Science of Computer Programming}

\begin{document}

\begin{frontmatter}

\title{\titulo}

\author[dcomp]{Fellipe A. Ugliara}
\ead{ugliara.fellipe@gmail.com}
\author[dcomp]{Gustavo M. D. Vieira\corref{cor1}}
\ead{gdvieira@ufscar.br}
\author[dcomp]{José de O. Guimarães}
\ead{jose@ufscar.br}
\cortext[cor1]{Corresponding author}

\address[dcomp]{DComp -- CCGT -- UFSCar, Sorocaba, Brazil}

\begin{abstract}
  Replication can be used to increase the availability of a service by
  creating  many  operational  copies  of its  data  called  replicas.
  Active  replication  is  a  form  of  replication  that  has  strong
  consistency semantics, which are easier to reason about and program.
  However, creating replicated services using active replication still
  demands  from the  programmer  the knowledge  of  subtleties of  the
  replication  mechanism.   In this  paper  we  show  how to  use  the
  metaprogramming infrastructure  of the  Cyan language to  shield the
  application programmer from these  details, allowing easier creation
  of   fault-tolerant    replicated   applications    through   simple
  annotations.
\end{abstract}

\begin{keyword}
  replication \sep metaprogramming \sep code generation
\end{keyword}

\end{frontmatter}

\section{Introduction}

Distributed computing offers the  promise of increased reliability and
performance  compared  to   traditional,  centralized  computing.   In
particular, greater reliability can  be achieved by \emph{replicating}
a service among many hosts to ensure availability of a service even in
the  presence  of faults.   Each  copy  of  the  service is  called  a
\emph{replica} and there are many strategies to create such replicated
service  that   usually  offer  a  balance   between  consistency  and
scalability~\cite{gray96, vogels2009}.  Among these techniques, a very
straightforward    and   studied    one    is   called    \emph{active
  replication}~\cite{schneider1990implementing}.

The  principle underlying  this technique  is to  consider the  system
being replicated as a deterministic state machine, which has its state
changed   only  by   well  defined   transitions.   Put   in  a   more
object-oriented way,  the system is modeled  by a set of  objects that
only  change  state  deterministically  by  calling  a  known  set  of
methods. To replicate the service, we have to identify each transition
before it happens, distribute the  information about the occurrence of
this  transition  and  its  data  to  all  replicas  and  execute  the
transition  in  all of  them.   Based  on  our assumption  that  these
transitions  are deterministic,  if we  are able  to distribute  these
transitions among  the replicas in  a strict order, the  replicas will
progress along the exactly same  states. These identical replicas will
be able  to provide the  required service in an  indistinguishable way
from each other.

To make the  task of creating a replicated  service easier, frameworks
such   as   Treplica~\cite{vieira08a,  vieira2010implementation}   and
OpenReplica~\cite{altinbuken2012commodifying}  were   created.   These
frameworks help  to create a replicated  system by taking care  of the
distribution, ordering  and execution  of the transitions  selected by
the application  programmer. The  integration of the  application into
these  frameworks happens  differently  depending  on the  programming
language  used. In  procedural  languages the  integration happens  by
function  calls to  the  framework and  callbacks  from the  framework
placed by the programmer. In object-oriented languages the integration
happens by  creating the classes  of the program by  extending classes
provided by the  framework.  Regardless of the  approach employed, the
linking  of   application  and   framework  usually   requires  adding
boilerplate code, intertwined with application code.

Current replication frameworks, albeit useful, only help with the
  communication and  ordering  of  transitions required  by
active replication. Other requirements  of this replication technique,
such as  a well defined set  of mutator methods and  the deterministic
nature of  these methods, are non  trivial and completely left  to the
application   programmer.   This   happens  because   the  traditional
procedural and object-oriented languages in which these frameworks are
built   are  not   suitable   to   enforce  these   nonfunc\-tion\-al
requirements.

Traditional languages lack mechanisms to  allow a program or framework
to change and validate its own code. Languages
that support metaprogramming~\cite{damavsevivcius2015taxonomy}
allow programs to inspect and modify their own code. Metaprogramming
has      been     used      to      translate        domain      specific
languages~\cite{rentschler2014designing},       implement       design
patterns~\cite{blewitt2005automatic}, perform  source code validations
at compile  time~\cite{chlipala2013bedrock}, to detect  defects in
object-oriented programs~\cite{mekruksavanich2012analytical}, produce code at runtime \cite{Goldberg:1983:SLI:273}, change the syntax of a language (as Lisp macros) \cite{Seibel:2012:PCL:2339396}, and implement pluggable type systems \cite{Markstrum:2010:JDP:1667048.1667049}.

In this paper we show how to use the metaprogramming infrastructure of
the Cyan  language~\cite{guimaraes2013cyan} to  transparently generate
and  validate  integration code  that  uses  the Treplica  replication
framework~\cite{vieira08a}.  We were able to use \emph{metaobjects} in
a centralized  object-oriented program to  isolate the set  of mutator
methods that change the state of a set of objects, and to generate the
appropriate extended classes to integrate with Treplica.  The approach
is    similar    in    essence    to    OpenMP~\cite{dagum1998openmp},
OpenACC~\cite{wienke2012openacc} and  other systems that  use compiler
directives   to   guide   the   automatic   generation   of   parallel
code.  However, our  approach is  easier  to use,  demanding just  the
addition of some annotations to methods and variable declarations, and it is
the  first  time metaobjects  are  used  to create  distributed  code.
Moreover, we were able to validate  the generated code with respect to
the  presence of  nondeterminism in  transitions by  flagging mutator
methods  that would  violate this  requirement.  Our  proposed set  of
metaobjects  is   able  to  replace  nondeterministic   methods  with
deterministic versions  and alert the  programmer if it still  finds a
call to a nondeterministic method inside mutator methods.

This  paper  is  structured  as  follows.   In  Section~\ref{cyan}  we
describe  the   Cyan  language  and   give  an  introduction   to  its
metaprogramming   features.    Section~\ref{treplica}  describes   the
organization of the  Treplica framework.  In Section~\ref{metaobjects}
we  describe the  proposed  metaobjects, how  to use  them  to turn  a
centralized  Cyan  program   into  a  replicated  one   and  how  they
work. Section~\ref{casestudy} shows the application of the metaobjects
to a more complex program as a demonstration of the feasibility of
our  approach.  The  paper  ends  with a  review  of  related work  in
Section~\ref{related}     and    some     concluding    remarks     in
Section~\ref{conclusion}. The  source code of the  Cyan compiler used,
the  set of  metaobjects  and  example applications  can  be found  at
\url{https://bitbucket.org/gdvieira/cyan.treplica.git}.

\section{The Cyan Language}
\label{cyan}

\subsection{Language Overview}

The language  used in  this paper is  Cyan~\cite{guimaraes2013cyan}, a
prototype-based  object-oriented  language.   Unlike
most  prototype-based languages,  Cyan  is statically  typed as  Omega
\cite{DBLP:books/daglib/0072762}, the language  it was initially based
on.  That makes  the  design of  Cyan  much closer  to  the design  of
class-based  languages  such  as  Java \cite{Gosling:2014:JLS:2636997},
C++ \cite{Stroustrup:2013:CPL:2543987}, or  C$\sharp$ \cite{csharp:2017:Online} than  to  other
prototype-based languages. Cyan programs  are compiled to produce Java
code, to be run in a Java virtual machine.

Prototypes  play  a  role  similar   to  classes.   Instead  of  using
\srcstyle{class} to declare a class, we use the keyword \srcstyle{object}
to  declare   a  prototype,  such  as   \srcstyle{Building}  shown  in
Figure~\ref{fig:Building}.  In this example, keyword \srcstyle{var} is
used to  declare a  field (instance  variable) and  \srcstyle{func} to
declare a  method.  In a field  declaration, the type comes  before the
field name  (\srcstyle{String} before \srcstyle{name} in  Line 17). Fields can only be private in Cyan, it is optional to use keyword \srcstyle{private} before a field declaration.
\srcstyle{self} refers  to the object  that received the  message. The
same  as  \srcstyle{self} in  Smalltalk \cite{Goldberg:1983:SLI:273} or  \srcstyle{this} in Java, C$\sharp$, or C++.

\begin{figure}[htbp]
\centering
\begin{lstlisting}[language=Java]
package main
object Building
    func init: String name,
               String address {
         self.name = name;
         self.address = address
    }
    func name:    String name
         address: String address {
         self.name = name;
         self.address = address
    }
    func getName -> String { return name }
    func getAddress -> String {
        return address
    }
    var String name
    var String address
end
\end{lstlisting}
\caption{A prototype in Cyan}
\label{fig:Building}
\end{figure}

Each  prototype  is  in  a  file with  its  own  name  (and  extension
\srcstyle{.cyan}). The  package declaration  should appear  before the
prototype  (Figure~\ref{fig:Building},  Line   1).   In  this  example
prototype  \srcstyle{Building}  is  in package  \srcstyle{main}.   For
conciseness, for  now on we  may show more  than one prototype  in the
same figure and without the package declaration.

A variable or field can  be declared using keywords \srcstyle{let} and
\srcstyle{var}. \srcstyle{let} is used to declare a read-only field or local variable
to which a value must be assigned.  For example, a variable
of type \srcstyle{Int} can be initialized as:
\begin{verbatim}
     let counter = 0;
\end{verbatim}
The variable  name is \srcstyle{counter} and  its type, \srcstyle{Int},
is deduced from the expression.   Variables and fields that can change
their  values  should  be  declared  with  keyword  \srcstyle{var}
(Figure~\ref{fig:Building}, Line 17).  Fields that are not preceded by
\srcstyle{var}    or   \srcstyle{let}    are   considered    read-only
(\srcstyle{let}) fields.

The syntax for message passing and method declaration is close to the Smalltalk syntax. Unary methods are those that do not take parameters, as \srcstyle{getName} of Line 13 of Figure~\ref{fig:Building}. Assuming \srcstyle{aBuilding}  is a variable
of type \srcstyle{Building},
\begin{verbatim}
    aBuilding getName
\end{verbatim}
is  the  sending of  the  \emph{unary  message} \srcstyle{getName}  to
object \srcstyle{aBuilding}.\footnote{More specifically, it  is the sending of
  message   \srcstyle{getName}   to   the  object   referred   to   by
  \srcstyle{aBuilding}.}     Messages    such     as    \srcstyle{-}    in
\verb|-counter| are also considered unary messages. In this example,
message \srcstyle{-} is being sent to object referred to by \srcstyle{counter}.

A keyword method is declared with identifiers ending with \srcstyle{:} each of which taking zero or more parameters as method \srcstyle{name:address:} of Lines 8-12 of Figure~\ref{fig:Building}. This method has two keywords. A keyword method may be called at runtime by keyword message passing, as in this example:
\begin{verbatim}
    aBuilding name: "Dahlia" address: "21 Drive";
\end{verbatim}
In this code, message \srcstyle{name: "Dahlia" address: "21 Drive"} is sent to the object \srcstyle{aBuilding}. If   \srcstyle{aBuilding}   refers   to  a   \srcstyle{Building}   object   at
runtime,\footnote{Even if  \srcstyle{aBuilding} has  type \srcstyle{Building},
it may refer to an object whose type is a   subprototype   of
  \srcstyle{Building}.} the  method called  would be that  declared in
line 8 of Figure~\ref{fig:Building}.

The name of a method may be an operator such as \srcstyle{+} or \srcstyle{<}. Method \srcstyle{+} should take no parameters (for unary \srcstyle{+}) or two parameters (for binary \srcstyle{+}). These methods are called as usual:  \srcstyle{1 + 2} is the sending of message \srcstyle{+ 2} to object \srcstyle{1}.

Object constructors are methods  with names \srcstyle{init} or \srcstyle{init:} (if there are parameters). They
cannot  be called  directly  by sending  messages.
For  each  method  \srcstyle{init}  or  \srcstyle{init:}  found  in  a
prototype,   the   compiler   creates  a   method   \srcstyle{new}   or
\srcstyle{new:} in the same prototype  with the same parameters as the
original method.   This \srcstyle{new} or \srcstyle{new:} method creates  an object and sends  to it
the  corresponding \srcstyle{init}  or \srcstyle{init:}  message.  For
example, the compiler adds a method\\
\verb|    |\srcstyle{new: String name, String address}\\
\noindent       to       prototype       \srcstyle{Building}       of
Figure~\ref{fig:Building}.\footnote{These  methods  are added  to  the
  compiler internal  representation, the  original source code  is not
  changed.} This method can only be called by sending a message to the
prototype itself:
\begin{verbatim}
   Building new: "Dahlia", "21 Drive"
\end{verbatim}
It  is a compile-time  error to
send a message  \srcstyle{new} or \srcstyle{new:} to  anything that is
not a prototype.

In prototype-based languages, prototypes are objects and they can be used in expressions. For example, \srcstyle{Int} is an object whose value is \srcstyle{0}.
\begin{verbatim}
   var Int two = (Int + 1)*2 + Int*Int;
   assert two == 2;
     //  ++ transforms both arguments to strings and
     //  concatenates them
   let String strZero = (Int asString) ++ 1;
   assert strZero == "01";
\end{verbatim}
However, In Cyan, only prototypes that declare an \srcstyle{init} method, a constructor without parameters, are considered objects with full rights.
The \srcstyle{init} method is used to build the object that is accessed by the prototype name (as \srcstyle{Int} in the expressions of the last example). This assures that the prototype fields are correctly initialized. That would not be the case with prototype \srcstyle{Building} of Figure~\ref{fig:Building}, which does not declare a method \srcstyle{init}. If the first message passing to \srcstyle{Building} is\\
\verb|    |\srcstyle{Building getName}\\
\noindent there would be an error: field \srcstyle{name} would not have been initialized.
A prototype without an \srcstyle{init} can receive some kinds of messages such as \srcstyle{new} and \srcstyle{new:}.

In Cyan, prototypes play a dual role: 1) they are types as with
classes in Smalltalk, Java, and C++, and 2) when used in expressions, they work like variables that refer to a fixed object of themselves.  For example, \srcstyle{Int}, when used inside an expression, refers to an object of prototype \srcstyle{Int} (itself) whose value is \srcstyle{0}.

Keyword \srcstyle{extends}  allows the inheritance of a superprototype by a subprototype. Inherited methods can be overridden in the subprototype, as usual. Java-like interfaces can be defined by using keyword ``\srcstyle{interface}'' instead of ``\srcstyle{object}'' when defining a prototype.

\subsection{The Cyan Metaobject Protocol}

Metaprogramming is a paradigm that allows programs to manipulate other
programs and  change themselves  in compilation  or execution time
\cite{10.1145/3354584} \cite{damavsevivcius2015taxonomy}.   Metaprogramming   has   a   broad
meaning, it encompasses any kind of program handling at compile-time, loading time (when the program is loaded into memory), and at runtime. In this  paper, we will limit ourselves to
transformations and  checks made  \textbf{at compile  time} by  a meta-level on a  base program. The program that is  changed or checked is
called  the {\it  base program} or simply {\it program}.  The code  that does  the changes  or
checks is called  {\it the meta level} or simply {\it metaprogram}.  The metaprogram can  be just a
set of classes  or functions and it acts as a  plugin to the
compiler, potentially changing how it parses, does type checking, generates
code, and so on.

In this paper, we consider that a \textit{metaobject  protocol (MOP)} is an interface between the  metaprogram, the program,
and the compiler.  It defines functions or methods  of the metaprogram
that should  be called when  a prototype is  inherited or a  method is
overridden in a  subprototype, when a field is accessed,  a message is
sent,  or  when   an  annotation  is  found  in  the   program  by  the
compiler. For  example, the MOP  defines that a  user-defined function
should be  called whenever a  prototype is inherited. Cyan  supports a
Metaobject Protocol, but not all languages that support metaprogramming do.

In Cyan,
the metaprogram consists of Java classes or Cyan prototypes. The compiler is written in Java, which makes it easy to write code in this language that interacts with the compiler. Since the Cyan compiler translates code to Java, the metaprogram can be written in Cyan too.

During the  compilation of  a program,  an \textit{annotation}  in the
source  code links  the program  (base level),  the
compiler, and the  metaprogram. An annotation is \srcstyle{@} followed by an identifier, optional parameters, and an optional DSL:
\begin{lstlisting}
  var Int sum = @eval("cyan.lang", "Int"){*
      var Int count = 0;
      for n in 1..10 {
        count = count + n
      }
      return count;
    *};
\end{lstlisting}
In this case, the optional DSL is given between \srcstyle{\{*} and \srcstyle{*\}}. It consists of Cyan code that is interpreted at compile-time. The result is the same as to replace the annotation by \srcstyle{55}. The metaprogram calls a Cyan interpreter at compile-time to produce the result. We will use just the identifier as the annotation name, as in ``annotation \srcstyle{eval}''.

\begin{figure}[htbp]
\centering
\begin{lstlisting}[language=Java]
object Person
    @init(name, age)
    func getName -> String { return name }
    func getAge -> Int { return age }
    String name
    Int age
end

object Program
    func run {
        let Person meg =
            Person new: "Meg", 3;
        let Person doki =
            Person new: "Doki", 5;
        meg getName println;
        doki getAge println;
        Out println: ("This is method " ++
           @compilationInfo(currentmethodname);
    }

end
\end{lstlisting}
\caption{Metaobjects in Cyan}
\label{fig:firstCyanExample}
\end{figure}

The example of Figure~\ref{fig:firstCyanExample} uses two annotations: \srcstyle{init} in Line 2 and \srcstyle{compilationInfo} in Line 18.
For each annotation, the compiler creates
a \textit{metaobject}  of a  \textit{metaobject class}. This  class is
written  in  Java,\footnote{Cyan can be used as the metaprogramming language. However, in this paper, we will consider that all metaobjects are made in Java, which is the language used to code all metaobjects cited in the text.}
as  the  compiler  is, and  is  associated  with a  Cyan
package.  When  the  package  is  imported  by  a  Cyan  program,  the
metaobject class is imported too. Then new behavior and new checks can be added to the code by importing packages and using annotations.
To simplify, we will
say  ``metaobject \srcstyle{init}''  and ``class  of \srcstyle{init}''
for the  metaobject associated  with annotation \srcstyle{init}  and the
metaobject class of metaobject \srcstyle{init}.

Package \srcstyle{cyan.lang}  is imported automatically by  every Cyan
program and this package keeps the metaobject classes of \srcstyle{init} and \srcstyle{compilationInfo}. Thus, these annotations can be used in any Cyan source code without explicitly importing a package, as is done in
the example of Figure~\ref{fig:firstCyanExample}. Then, the scheme the Cyan compiler uses for creating metaobjects is: a) when parsing the code and an annotation is found, look for a metaobject class in the imported packages. Method \srcstyle{getName} of the metaobject class should always return the annotation name; b) if there is no metaobject class, issue an error. Otherwise, create an object of the metaobject class and associates it with the annotation. For each annotation there is a unique metaobject and vice-versa.

During compilation, the Cyan compiler calls methods of the metaobjects associated with the annotations. They can insert code and do checks in the code based on information supplied by the compiler. In this example, metaobject \srcstyle{init} generates a constructor for \srcstyle{Person}. Figure~\ref{fig:firstExampleAfterInsertion} shows the resulting
\srcstyle{Person} prototype.

The constructor code is returned by the metaobject method as a string (in fact, a \srcstyle{StringBuilder} object) and inserted in the source code by the compiler. Only the source code in memory is changed, the original
file   with  prototype   \srcstyle{Person}  is   not  changed.
Annotation \srcstyle{compilationInfo} produces the current method name, \verb|"run"|.

Since the metaobject inserted an \srcstyle{init:} method in the prototype,  method \srcstyle{run}
of \srcstyle{Program}  can create  objects of \srcstyle{Person} using
the \srcstyle{new:} method (constructor).
Note that the Java class of metaobject \srcstyle{init} knows the types of fields \srcstyle{name} and \srcstyle{age}.
These types are necessary to generate the constructor. This information is supplied to the metaobject by the compiler.

\begin{figure}[htbp]
\centering
\begin{lstlisting}[language=Java]
object Person
    @init(name, age)
    func init: String name,
               Int age {
        self.name = name;
        self.age = age;
    }
    func getName -> String { return name }
    func getAge -> Int { return age }
    String name
    Int age
end
\end{lstlisting}
\caption{\textit{Person} generated during compilation}
\label{fig:firstExampleAfterInsertion}
\end{figure}

\begin{figure}[htbp]
    \centering
    \fbox{\includegraphics[width=12cm]{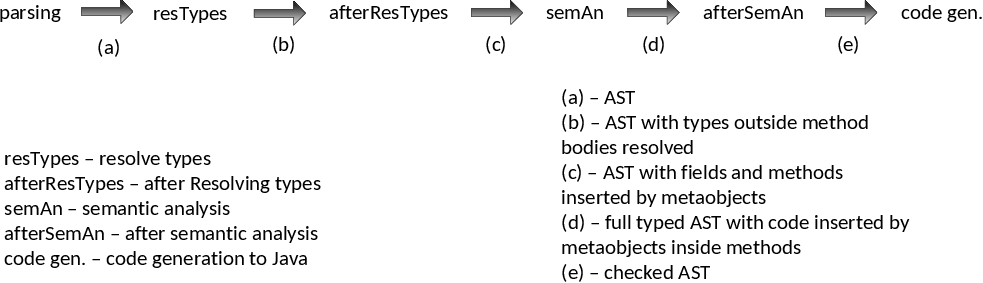}}
    \caption{The Cyan Compilation Phases}
    \label{fig:phases}
\end{figure}

In the remaining of this section,  we will give a simplified overview of the Cyan MOP. The Metaobject Protocol can only be understood by studying the compilation phases of the Cyan compiler, shown in Figure~\ref{fig:phases}. The compilation phases are parsing, resTypes (resolve types), afterResTypes (after resolving types), semAn (semantic analysis), afterSemAn (after semantic analysis), and code generation. The arrows indicate the data passed from one phase to the next. The data is described in the right-hand side of the Figure by the labels. Then, the arrow from parsing to phase resTypes is labeled with \textit{(a)}, which represents the Abstract Syntax Tree (AST). Parsing uses only local information, available in the current source code. Therefore, after parsing there is no type information in the AST. For example, for a method parameter that has type \srcstyle{Int}, the AST keeps only the string ``\srcstyle{Int}''. In phase resTypes, the compiler looks for a prototype \srcstyle{Int} and assigns a reference to its AST to a field \srcstyle{type} of the AST object that represents the method parameter. All references to types outside method statements are resolved in phase resTypes.

In phase afterResTypes, metaobjects can add fields and methods to the prototype in which their annotations are. The metaobject associated with an annotation that is in a prototype cannot add code to another prototype. A method of metaobject \srcstyle{init} is called in phase afterResTypes. It returns, as a string\footnote{As a \srcstyle{StringBuffer} object, in fact.}, the code to be added, an \srcstyle{init:} method. A copy of the source code, in memory, is changed and therefore the compilation starts again from phase parsing (but only for this file). Metaobject methods called in phase afterResTypes have access to
the AST of the prototype in which they are, except method statements. Then, metaobject methods called in this phase know the prototype name, superprototype, implemented interfaces, the name and type of fields, and method signatures.\footnote{A signature of a method is composed of its name, name and type of parameters, and return value type.} Metaobjects can then use this information to generate code, as does \srcstyle{init}.

Phase semAn is called ``the semantic analysis'' even though it only finishes this analysis, started in phase resTypes. In phase parsing, types are assigned to expressions inside methods. Every expression object of the AST has a ``\srcstyle{type}'' field that is \srcstyle{null} till the start of phase semAn. After this phase all ``\srcstyle{type}'' fields are resolved. Metaobjects can generate code inside methods only in this phase. This is what metaobject \srcstyle{compilationInfo} does in the example of Figure~\ref{fig:firstCyanExample}. It produces the current method name as a string.

After phase semAn, metaobjects cannot change the code anymore. But they can do checks in the code in phase afterSemAn. This phase is ideal for checks because there will be no further changes that can invalidate the checks. Currently, metaobjects cannot act in the last phase, code generation.

Metaobject methods can be called in phases parsing, afterResTypes, semAn, and afterSemAn.\footnote{Call to metaobject methods in phase parsing will soon be deprecated.} But how the compiler knows which method it should call in each phase? The answer is in the interfaces implemented by the classes of the metaobjects. The Cyan MOP provides several Java (and Cyan) interfaces that should be implemented by metaobjects. Each interface is associated with a compilation phase. Whenever a metaobject class implements an interface of phase afterResTypes, for example, all metaobject methods that override the interface methods are called in phase afterResTypes. As a concrete example, the class of metaobject \srcstyle{init} is \srcstyle{CyanMetaobjectInit} that inherits from class \srcstyle{CyanMetaobjectAtAnnot} (as any metaobject class of this paper) and implements interface \srcstyle{IAction\_afterResTypes}. This interface declares a method \srcstyle{afterResTypes\_codeToAdd} that is overridden by \srcstyle{CyanMetaobjectInit}. This method is called in phase afterResTypes by the compiler.

During the parsing  of prototype \srcstyle{Person} of Figure~\ref{fig:firstCyanExample}, the compiler
creates an object of class \srcstyle{CyanMetaobjectInit} when it finds the annotation \srcstyle{init} in line~2. The name ``\srcstyle{init}'' is associated with this class because the compiled form of it, a Java ``\srcstyle{.class}'' file, is in a directory of package \srcstyle{cyan.lang} automatically imported by every Cyan source file. And method \srcstyle{getName()} of an object of this class returns \srcstyle{"init"}. The compiler knows how to associate an annotation to a metaobject class.

The \srcstyle{CyanMetaobjectInit} object created in phase parsing is the \textit{metaobject} associated with annotation \srcstyle{init}. Method \srcstyle{afterResTypes\_codeToAdd} of this metaobject is called in phase afterResTypes to produce the Cyan \srcstyle{init:} method, which is then inserted in the code by the compiler. All methods of interface \srcstyle{IAction\_afterResTypes} are called on the metaobject. The other methods, in this case, do nothing.

The class of metaobject \srcstyle{compilationInfo} implements interface \srcstyle{IAction\_semAn} which declares a single method, \srcstyle{semAn\_codeToAdd}. During the semantic analysis, phase semAn, the compiler calls method \srcstyle{semAn\_codeToAdd} of the metaobject associated with annotation \srcstyle{compilationInfo} of line 18 of Figure~\ref{fig:firstCyanExample}. This method produces the string \srcstyle{"run"} which is inserted in the code by the compiler.

Note that there is a one-to-one relationship between annotations and metaobjects. Two \srcstyle{compilationInfo} annotations, even with the same parameters, will cause the creation of two metaobjects in phase parsing.
Annotations can be attached to declarations such as prototypes, methods, local variables, packages, and the program. Or they can be \textit{free} as \srcstyle{compilationInfo}, which is not attached to anything.

There are several \textit{other} interfaces associated with phases parsing, afterResTypes, semAn, and afterSemAn. They are used to:
\begin{enumerate}
  \item parse the DSL attached to the annotation (parsing);
  \item create new prototypes (parsing, afterResTypes, semAn);
  \item intercept field access (semAn);
  \item intercept message passings (semAn);
  \item create code after a local variable declaration (semAn);
  \item simulate the existence of fields (semAn);
  \item simulate the existence of methods (semAn);
  \item intercept the overridding of a method (afterSemAn);
  \item intercept the inheritance of a prototype (afterSemAn);
  \item check any kind of declaration (afterSemAn);
  \item check message passing (afterSemAn).
\end{enumerate}

Metaobjects are used, in this paper, for replication. We will show the details of metaobject implementation later on using the replication metaobjects.

\section{Treplica}
\label{treplica}

Treplica~\cite{vieira08a} is a framework written in Java that provides
an  active replication  structure  for the  development of  replicated
distributed applications.  In this section  we briefly describe how to
program  this  framework through  a  binding  developed for  the  Cyan
language  with the  purpose of  characterizing the  programming effort
required to program a replicated  application using the framework. The
source   code  of   this  binding   can  be   found  in   the  project
tree\footnote{\url{https://bitbucket.org/gdvieira/cyan.treplica.git}}
under \srcstyle{lib/treplica}.

A  complete  description of  Treplica  is  beyond  the scope  of  this
paper. A  full description  of Treplica,  including its  original Java
programming          interface,          can         be          found
in~\cite{vieira2010implementation}.   Nonetheless,   we  will  briefly
summarize  two   properties  of   Treplica  that  are   important  for
understanding how to use the  framework: its consistency guarantee and
the way active replication works.

Replicated applications can be classified  in two types reflecting the
way updates  are propagated:  eager and lazy~\cite{gray96}.   Assume a
client is  contacting a  replicated server to  execute a  service, and
that the  execution of this service  changes the state of  the server.
Eager replication propagates  and ensures that all  changes are stable
in  all replicas  before  returning to  the  client. Lazy  replication
executes  and  returns  to  the client  immediately,  propagating  the
changes at  a later, more  convenient time. Thus, in  lazy replication
the replicas can diverge immediately after an update is made and later
converge  back to  a  shared  state, while  in  eager replication  the
replicas never diverge.

Lazy replication has  the advantage that it  exchanges fewer messages,
resulting  in lower  consumption of  system resources  and less  delay
associated  with  message exchange.   This  is  a consequence  of  the
flexibility  to decide  when  to propagate  changes. Lazy  replication
protocols are therefore faster than eager protocols that must exchange
messages before a change is made to a replica~\cite{gray96}.  However,
to  achieve  this  higher  efficiency  lazy  replication  relaxes  the
consistency  of  the   changes  made  to  the   replicas,  leading  to
inconsistencies   and   requiring   conciliation  of   the   resulting
data~\cite{gray96}.

For  example, suppose  a bank  system that  uses lazy  replication and
stores a checking account with \$100 in it.  Two independent purchases
of \$100  can be made concurrently  in different replicas of  the bank
system.  At this moment the  system is inconsistent, because the owner
of the account  has spent more than was available.   But, eventually a
process of reconciliation  will happen and, for example,  one of the
purchases will be  canceled for lack of funds. From  the point of view
of the system this is simple, but  it may be a nuisance to both client
and merchant. Worse still, if the goods were delivered, it may even be
impossible  to  revert  such   transactions.   Thus,  the  application
programmer has  to be aware  of the  need for reconciliation  in every
transaction, making the application more complicated.

Eager replication  is more  costly, but  it does  not suffer  from the
problem created by divergence and reconciliation of data. Data changes
in  a eager  replication  protocol usually  guarantee  a criterion  of
consistency called  one-copy-serializability~\cite{bernstein87}.  This
criterion  specifies that  concurrent  changes  to replicated  objects
appear as a  series of changes over a single  logical copy.  Solutions
that  use  this   type  of  consistency  bear   more  similarities  to
centralized   systems   and   are   easier   to   reason   about   and
program~\cite{schneider1990implementing}.  However, even in this case,
the  programming   of  these   applications  still   poses  significant
challenges~\cite{burrows2006chubby}.

Active replication  is a type  of eager replication that  ensures that
many  copies of  a single  application known  as \emph{replicas}  keep
their state up to date and consistent even as changes are made as part
of the  application operation.  It  works by assuming  the application
behaves like  a deterministic  state machine,  which only  changes its
internal state by deterministically  executing transitions.  If one is
able  to  execute the  same  transitions  in  the  same order  in  all
replicas, they  will end up with  the same resulting state  due to the
deterministic nature of  the transitions. As a  consequence, an active
replication  framework must  provide a  reliable way  of disseminating
transitions in a ordered way among  all replicas and it should provide
a  programming interface  that allows  regular applications  to behave
like deterministic state  machines even if they are  not programmed as
such.

Treplica solves the problem of  disseminating transitions by using the
Paxos algorithm  to ensure the  transitions reach all replicas  in the
same order,  even in the presence  of failures~\cite{lamport2006fast}.
Treplica  solves the  programming  interface problem  by providing  an
object-oriented abstraction that  defines the very simple  notion of a
shared state  and well defined  changes to this state.   The resulting
programming  interface  is  as   close  as  possible  to  conventional
centralized  applications~\cite{vieira08a},   making  the  replication
\emph{mechanism} transparent to the developers.

The shared state of an application is defined by its \emph{context}, a
single  object that  stores the  application  data in  its fields.  In the  Treplica framework this object  should extend the
\srcstyle{Context}  prototype  and   be  serializable.\footnote{In Cyan, every prototype is serializable.}  Serialization
allows  an  object to  be  transformed  to text,  transmitted  between
different  hosts and  transformed  back  to a  clone  of the  original
object.     Figure~\ref{fig:TreplicaDados}    shows   the    prototype
\srcstyle{Info}, an example of  an application context. This prototype
contains  two variables:  an \srcstyle{Int}  and a  \srcstyle{String},
representing the state of this application.

\begin{figure}[htbp]
\centering
\begin{lstlisting}[language=Java]
object Info extends Context {
    var String text
    var Int number

    func setNumber: Int number {
        self.number = number;
    }

    func setText: String text {
        self.text = text;
    }

    func getText -> String {
        return self.text;
    }
}
\end{lstlisting}
\caption{Prototype to be replicated}
\label{fig:TreplicaDados}
\end{figure}

The prototype  \srcstyle{Info} has two \srcstyle{set}  methods used to
assign values to  its private variables.  More  importantly, these two
methods  allow changing  the state  of the  application context.   The
Treplica framework  considers a message  sent to an object  that calls
one of  these mutator methods  to be  equivalent to a  transition that
changes the context  state in a deterministic way.  The framework then
defines  a   way  to  capture   and  represent  this  message   as  an
\emph{action}.

The prototypes that extend prototype \srcstyle{Action} implement these
actions.  They contain  as fields the parameters of the  message to be
sent and are  serializable, allowing the record of this  message to be
sent  to   other  hosts.   Also,   they  must  implement   the  method
\srcstyle{executeOn:} that defines  the keywords of the  message to be
sent by encoding an actual message  passing operation using the fields as
parameters.    The   target   object    of   the   message   sent   by
\srcstyle{executeOn:}  is  defined by  a  parameter  received by  this
method.     Figure~\ref{fig:TreplicaAction}   shows    the   prototype
\srcstyle{SetTextAction},  which implements  a  transition that  sends
message \srcstyle{setText:} to an \srcstyle{Info} object. Statement \srcstyle{type-case} is a safe way of downcasting in Cyan. \srcstyle{type} may be followed by one or more \srcstyle{case} clauses. The statement tries to cast the argument to \srcstyle{type} to the type that appears after \srcstyle{case}. In this example, the statement tries to cast \srcstyle{context} to \srcstyle{Info}. If this succeeds, the value is put in variable \srcstyle{info} and the \srcstyle{case} statements are executed.

\begin{figure}[htbp]
\centering
\begin{lstlisting}[language=Java]
object SetTextAction extends Action {
    var String updateText

    func init: String text {
        self.updateText = text;
    }

    override
    func executeOn: Context context {
        type context
            case Info info {
                info setText: self.updateText;
            }
    }
}
\end{lstlisting}
\caption{Prototype that implements a transition}
\label{fig:TreplicaAction}
\end{figure}

The actual firing of a transition is implemented by Treplica. When the
application  wants to  change  its state,  it  creates an  appropriate
action  object and  passes  it to  Treplica  in a  \srcstyle{execute:}
message  sent  to  an  object of  prototype  \srcstyle{Treplica}.   The
framework then  sends a  copy of  this object  to the  other replicas,
properly    ordered,   and    all   of    them   send    the   message
\srcstyle{executeOn:} to the received object,  passing a local copy of
the context as  the parameter.  Therefore, all the copies  will end up
with contexts with the same values.

For this to  work, no changes to the context  can happen without being
represented  as  actions  and  the  actions  passed  to  Treplica  for
execution.   The  application  places  its state  under  care  of  the
framework    during   the    application   initialization,    when   a
\srcstyle{Treplica} object  is instantiated.  In Cyan,  that is usually
done  in  a  method  called  \srcstyle{run:} in  a  prototype  called
\srcstyle{Program}.\footnote{This is the default that can be changed in the \textit{project file}, a file that declares the packages and other configurations of a program.}
Figure~\ref{fig:TreplicaMain}   shows   how   a
Treplica object is  declared and initialized in each  replica. This is
also   an    example   of   how    an   object   of    the   prototype
\srcstyle{SetTextAction}  is  passed  as  an argument  to  the  method
\srcstyle{execute:} of the Treplica object.

\begin{figure}[htbp]
\centering
\begin{lstlisting}[language=Java]
object Program {
    func run: Array<String> args {
        let info = Info new;
        let treplica = Treplica new;
        treplica runMachine: info
                 numberProcess: 3
                 rtt: 200
                 path: "/var/tmp/magic" ++ args[0];

        let action = SetTextAction new: "text";
        treplica execute: action;
    }
}
\end{lstlisting}
\caption{Treplica configuration and execution}
\label{fig:TreplicaMain}
\end{figure}

The sequence diagram of Figure~\ref{fig:trepexecexamplecyan} shows the
execution flow of the  example in Figure~\ref{fig:TreplicaMain}.  Both
Replicas  A and  B start  execution  in the  \srcstyle{run} method  of
prototype   \srcstyle{Program}.   The   context  of   the  application
(\srcstyle{Info})  and the  Treplica object  (\srcstyle{Treplica}) are
initialized during the  execution of this method.  Replica  A wants to
change  the replicated  state,  so it  creates  an appropriate  action
object  (\srcstyle{SetTextAction})  and sends  an  \srcstyle{execute:}
message to Treplica with the  object as parameter. Treplica will order
and distribute the action object to  both replicas and it will execute
the action on both, independently, by sending an
\srcstyle{executeOn:} message to the local action object.

\begin{figure}[htbp]
  \centering
  \tikzset{every picture/.append style={transform shape,scale=0.75}}
  \begin{sequencediagram}[font=\fontsize{0.32cm}{0.35cm}\selectfont\ttfamily]
    \draw (5.6,-1)  node[right] {\small Replica A};
    \draw (8.3,-1)  node[right] {\small Replica B};
    \draw (8,0) -- (8,-4.8);
    \draw (8,-5.8) -- (8,-10.3);
    \newthread{InstAMain}{Program}
    \newinst{InstAInfo}{Info}
    \newinst{InstAUpdate}{SetTextAction}
    \newinst{InstAState}{Treplica}

    \newinst[0.6]{InstBState}{Treplica}
    \newinst{InstBUpdate}{SetTextAction}
    \newinst{InstBInfo}{Info}
    \newthread{InstBMain}{Program}

    \begin{call}{InstAMain}{Info new}{InstAInfo}{info} \end{call}
    \begin{call}{InstAMain}{Treplica new}{InstAState}{treplica} \end{call}
    \begin{call}{InstAMain}{runMachine:}{InstAState}{} \end{call}

    \prelevel \prelevel \prelevel \prelevel \prelevel \prelevel

    \begin{call}{InstBMain}{Info new}{InstBInfo}{info} \end{call}
    \begin{call}{InstBMain}{Treplica new}{InstBState}{treplica} \end{call}
    \begin{call}{InstBMain}{runMachine:}{InstBState}{} \end{call}

    \begin{call}{InstAMain}{execute:}{InstAState}{}
      \postlevel
      \begin{call}{InstAState}{executeOn:}{InstAUpdate}{}
        \begin{call}{InstAUpdate}{setText:}{InstAInfo}{} \end{call}
      \end{call}

      \prelevel \prelevel \prelevel
      \prelevel \prelevel

      \begin{messcall}{InstAState}{<Paxos>}{InstBState}{}
        \postlevel
        \begin{call}{InstBState}{executeOn:}{InstBUpdate}{}
          \begin{call}{InstBUpdate}{setText:}{InstBInfo}{} \end{call}
        \end{call}
      \end{messcall}
    \end{call}
  \end{sequencediagram}
  \caption{Treplica execution sequence diagram}
  \label{fig:trepexecexamplecyan}
\end{figure}

The  way  Treplica   encodes  message  passing  is   very  simple  and
straightforward, but it requires  the application programmer to create
much  boilerplate  code in  the  form  of action  objects.   Moreover,
isolating message passing  isn't enough to achieve  replication, it is
necessary  that the  method activated  by the  message doesn't  create
external effects  besides changing  the context  state and  that these
changes are deterministic. It is  the responsibility of the programmer
to be aware of these requirements and avoid breaking them.

For example,  in Figure~\ref{cod:PropostaConsistencia} we  introduce a
small  change to  the  \srcstyle{setText:}  method of  \srcstyle{Info}
prototype to  make it nondeterministic.   The problem brought  by the
change is that every time the \srcstyle{setText:} method is called the
value  set  will   be  different,  even  if  the   starting  state  of
\srcstyle{Info}  and the  parameters  of  \srcstyle{setText:} are  the
same.   This will  make  the replicas  diverge,  as the  deterministic
behavior of  the transition will be  violated. Later in the  paper, we
show how this problem can be detected and solved.

\begin{figure}[htbp]
\centering
\begin{lstlisting}[language=Java]
...
object Info extends Context {
  ...
  func setText: String text {
    var date = System currentTime asString;
    self.text = text ++ date;
  }
  ...
}
\end{lstlisting}
\caption{Nondeterministic \srcstyle{setText:} method}
\label{cod:PropostaConsistencia}
\end{figure}

\section{Metaobjects for Replication}
\label{metaobjects}

\subsection{Overview and Use}

The creation of  a replicated application using  Treplica requires the
construction of  a prototype representing the  application context and
as many actions as messages this  context can receive. For each action
it is necessary to define a  new prototype with the correct number and
type of  fields, besides  sending the  correct message  to the application
context when asked by Treplica. We have shown in the last section this
task isn't hard, but it  requires the tedious and error-prone creation
of a lot of boilerplate code. Now we are going to show how programming
of  replicated  application can  be  simplified  by  the use  of  Cyan
metaobjects. We
first describe how  to use the metaobjects and in  the next section we
describe how the metaobjects are created and how they work.

A good  programming practice when  using Treplica is to  create action
prototypes that do not have  application functional behavior and limit
themselves  to send  a  single message  with  the correct  parameters.
Although  Treplica does  not  impose these  restrictions, by  creating
data-only action  prototypes the coupling between  the application and
Treplica  is   reduced.   Loosely   coupled  modules  are   easier  to
reuse~\cite{eder94}  and  align   better  with   replication  as   a
nonfunctional           requirement.           For           example,
Figure~\ref{fig:TreplicaAction}        shows         the        action
\srcstyle{SetTextAction}  that  represents   the  sending  of  message
\srcstyle{setText:}  in the  form of  a prototype.   Notice that  this
prototype,  besides its  \srcstyle{executeOn:}  method,  has a  simple
constructor that initializes the fields of the created object with the
same parameters used  afterwards to send the  encoded message. Because
of  this  regularity,   the  creation  of  these   prototypes  can  be
standardized. The metaobject \srcstyle{treplicaAction} will create
an appropriate action prototype when its annotation is attached to a method declaration
of the application context.

For      example,      the      \srcstyle{Info}      prototype      in
Figure~\ref{fig:TreplicaDadosMeta} is  similar to the one  depicted in
Figure~\ref{fig:TreplicaDados},         except        that         the
\srcstyle{treplicaAction}  annotation   is  attached  to   the  method
\srcstyle{setText:}.  The  metaobject associated with  this annotation
modifies the prototype \srcstyle{Info}, adding  a new method to it and
creating  a  new prototype  that  represents  the sending  of  message
\srcstyle{setText:}     as    a     Treplica    action.      Prototype
\srcstyle{SetTextAction}  of  Figure~\ref{fig:TreplicaAction}  is  not
necessary  any more,  since  the metaobject  \srcstyle{treplicaAction}
adds  an  equivalent  prototype  to the  program  during  compilation.
Moreover,   \srcstyle{setText:}   messages   sent  directly   to   the
application  context  will  be  ``intercepted'' and  replaced  by  the
creation of a  suitable action object and the sending  of this object
to  Treplica  as an  \srcstyle{execute:}  message  for replication  and
execution by the framework.

Metaobject \srcstyle{treplicaAction} replaces nondeterministic message passings inside the annotated method by deterministic message passings. A list of replacements is given in files loaded at the start of the compilation --- the details of how this is made will be given later on.

\begin{figure}[htbp]
\centering
\begin{lstlisting}[language=Java]
package main
import treplica
object Info extends Context {
    var String text
    ...
    @treplicaAction
    func setText: String text {
        self.text = text;
    }
    ...
}
\end{lstlisting}
\caption{Replicated prototype using metaobjects}
\label{fig:TreplicaDadosMeta}
\end{figure}

Initialization  of  the  application   context  and  of  the  Treplica
framework   also  happen   in   a  standardized   way   as  shown   in
Figure~\ref{fig:TreplicaMain}.
This initialization is automated by the \srcstyle{treplicaInit} metaobject, whose annotations
can be attached to variable declarations whose types are
subprototypes of \srcstyle{Context}
This metaobject  has a  double
function: it marks  a variable as holding the  application context and
it  initializes  Treplica.   The  explicit indication  of  the  object
holding the application context is  important because only the methods
belonging  to   the  context   prototype  can   be  marked   with  the
\srcstyle{treplicaAction} metaobject.

For     example,     in     Figure~\ref{fig:TreplicaMainMeta}     the
\srcstyle{treplicaInit}   metaobject    is   attached    to   variable
\srcstyle{info}, with the parameters of the desired Treplica instance.
This metaobject  adds code after the variable declaration, during compilation,
to create a  new instance of \srcstyle{Treplica} and assign  to it the
object  \srcstyle{info},  similar  to  the  method  \srcstyle{run}  in
Figure~\ref{fig:TreplicaMain}. Note that the annotation \srcstyle{treplicaInit} can be attached to the declaration because the type of \srcstyle{info}, \srcstyle{Info}, is a subprototype of \srcstyle{Context}.

\begin{figure}[htbp]
\centering
\begin{lstlisting}[language=Java]
object Program {
    func run: Array<String> args {
        var local = "/var/tmp/magic" ++ args[0];
        @treplicaInit( 3, 200, local )
        var info = Info new;
        info setText: "text";
    }
}
\end{lstlisting}
\caption{Treplica configuration using metaobjects}
\label{fig:TreplicaMainMeta}
\end{figure}

Using     these    two     metaobjects     only     the    code     in
Figures~\ref{fig:TreplicaDadosMeta}   and   \ref{fig:TreplicaMainMeta}
need to be written.  This code is more compact, easier  to read, and is
independent from  any replication concerns or  implementation details.
This way we can avoid some  of the pitfalls created by the programming
interface of Treplica and create less complex applications.

\subsection{Implementation of the Metaobjects}

This section  shows how the metaobjects  \srcstyle{treplicaAction} and
\srcstyle{treplicaInit} are  implemented. We describe  the interface of metaobjects
 with the Cyan MOP in  Java.
The  source code  of  these metaobjects  can be  found  in the  project
tree\footnote{\url{https://bitbucket.org/gdvieira/cyan.treplica.git}}
under  \srcstyle{proj/meta/treplica}.  The  compiled  version of  both
classes, the  ``.class'' file, are put  in directory ``\verb|--meta|''
of package \srcstyle{treplica}.  When this package is  imported, as in
Figure~\ref{fig:TreplicaDadosMeta}, the associated  annotations can be
used.

\begin{figure}[htbp]
\centering
\begin{lstlisting}[language=Java]
object Info extends Context {
   ...
   func setText: String text {
       var action = InfosetText new: text;
       self getTreplica execute: action;
   }
   @treplicaAction
   func setTextTreplicaAction:
          String text {
       self.text = text;
   }
   ...
}
\end{lstlisting}
\caption{Prototype \srcstyle{Info} modified}
\label{fig:InfoChange}
\end{figure}

\srcstyle{CyanMetaobjectTreplicaAction} is the Java class implementing
the  metaobject  \srcstyle{treplicaAction}. This  association  happens
because its method \srcstyle{getName()} returns \verb|"treplicaAction"|. This class implements several interfaces, which are described below together with their role in the metaobject.
\begin{enumerate}[(a)]
\item \srcstyle{IAction\_afterResTypes} from which methods \srcstyle{afterResTypes\_codeToAdd} and \srcstyle{afterResTypes\_renameMethod} are defined. Method 
    \srcstyle{afterResTypes\_codeToAdd}
    adds a new method with the same name as the annotated method. In the example of Figure~\ref{fig:InfoChange}, the metaobject of Figure~\ref{fig:TreplicaDadosMeta} adds method \srcstyle{setText:}. The code of \srcstyle{setText:} is produced as a string based on the data of the original method. The needed information is the method name, the parameter names and types, and return value type. The first line of method \srcstyle{afterResTypes\_codeToAdd} is\\
    \verb|  WrMethodDec md = (WrMethodDec) this.getAttachedDeclaration();|\\
    \noindent \srcstyle{md} refers to the object of the AST that represents the method annotated  with \srcstyle{treplicaAction}, \srcstyle{setText:}. Using \srcstyle{md}, we can get the method name, list of parameters, and return value type with the following method calls:
\begin{lstlisting}
md.getName()
md.getMethodSignature().getParameterList()
md.getMethodSignature().getReturnTypeExpr().getType()
\end{lstlisting}
This is all the information needed to create a new \srcstyle{setText:} method in string format, which is returned by method \srcstyle{afterResTypes\_codeToAdd}.

Method \srcstyle{afterResTypes\_renameMethod}  renames the original annotated method. In the example, \srcstyle{setText:} is renamed to \srcstyle{setTextTreplicaAction:}. The information needed for this is obtained as in method \srcstyle{afterResTypes\_codeToAdd}.

\item    \srcstyle{IActionNewPrototypes\_afterResTypes}   from    which   method
  \srcstyle{afterResTypes\_NewPrototypeList}
   \noindent is redefined  for  creating a  new
  prototype implementing the Treplica action. In the example, it is prototype
    \srcstyle{InfosetText} of Figure~\ref{fig:TreplicaMetaActionResult}. The new prototype is created as a string. The information needed for that is got as in method \srcstyle{afterResTypes\_codeToAdd}.

\item       \srcstyle{IAction\_semAn}       from      which       method
  \srcstyle{semAn\_codeToAdd} replaces  calls to non\-deter\-mi\-nis\-tic
  methods by calls to deterministic ones. The list of nondeterministic methods and their replacements are put in files loaded before the compilation starts. We will describe this nondeterminism
    removal operation in more detail in the next section.

\begin{figure}[htbp]
\centering
\begin{lstlisting}[language=Java]
object InfosetText extends Action {
    var String textVar
    func init:  String text {
        self.textVar = text;
    }

    override
    func executeOn: Context context {
        type context
            case Info obj {
                obj setTextTreplicaAction:  textVar;
            }
    }
}
\end{lstlisting}
\caption{Prototype created by \srcstyle{treplicaAction}}
\label{fig:TreplicaMetaActionResult}
\end{figure}

\item \srcstyle{ICheckDeclaration\_afterSemAn} from which method 
\srcstyle{afterSemAn\_checkDeclaration} 
 is redefined to check if there are any calls to nondeterministic methods in the final code. This should not be necessary since method \srcstyle{semAn\_codeToAdd} replaces all possible calls to nondeterministic methods by calls to deterministic ones. However,  other metaobjects may have introduced nondeterministic method calls in phase semAn (if they add code, this code will not be visible to \srcstyle{treplicaAction} in this phase).

\end{enumerate}

Class   \srcstyle{CyanMetaobjectTreplicaInit}   is  the   Java   class
implementing    metaobject   \srcstyle{treplicaInit}.     This   class
implements  interface
 \srcstyle{IActionVariableDeclaration\_semAn}  
 and redefines its method  \srcstyle{semAn\_codeToAddAfter}. This method adds
code  after the  variable  declaration to  create  and initialize  the
\srcstyle{Treplica} object.   For example,  this metaobject  takes the
annotated  code  in  Figure~\ref{fig:TreplicaMainMeta}  and  adds  the
initialization  code in  Lines  5--10 of  Figure~\ref{fig:InitChange}.
The    newly    added    code    makes    \srcstyle{info}    reference
\srcstyle{treplicainfo}, the treplica object, and vice-versa.

\begin{figure}[htbp]
\centering
\begin{lstlisting}[language=Java]
object Program {
    func run: Array<String> args {
        var local = "/var/tmp/magic" ++ args[0];
        var info = Info new;
        var treplicainfo = Treplica new;
        treplicainfo runMachine: info
                     numberProcess: 3
                     rtt: 200
                     path: local;
        info setTreplica: treplicainfo;
        info setText: "text";
    }
}
\end{lstlisting}
\caption{Prototype \srcstyle{Program} modified}
\label{fig:InitChange}
\end{figure}

\subsection{Nondeterminism Detection}

Besides  removing  boilerplate  code  from  a  program,  the  proposed
metaobjects offer  initial support for validating  if the resulting
code  is  indeed  able  to  be   replicated.   As  we  have  shown  in
Section~\ref{treplica}, methods  that change the state  of the context
must be deterministic and should  not create external effects. In this
work, we considered only the question of identifying nondeterministic
methods and optionally replacing  them with deterministic versions. In
the  previous  section  we  have briefly  cited  these  tests  and
substitutions and now we describe them in a bit more depth.

Identifying deterministic  methods is complex  and this work  does not
offer comprehensive  solutions to this problem.   These methods are usually  operating system services that aren't
deterministic  by nature,  such as  reading  the time  or receiving  a
packet  from the  network.  Nondeterministic  methods and deterministic replacements for them are  defined  by the  developer  in \textit{rule
files}.
Based on the information of these \textit{rule files}, metaobject \srcstyle{treplicaAction} can replace nondeterministic calls in the annotated method by deterministic method calls.

Rule files are loaded in memory by metaobject \srcstyle{loadNonDeterministicFiles} whose annotation should be attached to the program in the \textit{project file}. Every Cyan program should have a \textit{project file} that describes its packages. The simplest example is just
\begin{lstlisting}[language=Java]
program
    package treplica at "C:\projeto\lib\treplica"
    package cyan.math at "C:\Cyan\lib\cyan\math"
    package main
    package other
\end{lstlisting}
The compiler will consider that the program is composed of packages \srcstyle{treplica} and \srcstyle{cyan.math}, in the directories given after ``\srcstyle{at}'', and \srcstyle{main} and \srcstyle{other}, that should be in the same directory as the project file. The files of a package in Cyan should be in a directory whose name is the package name.

Annotations can be attached to the program and to packages:
\begin{lstlisting}[language=Java]
import treplica at "C:\projeto\lib\treplica"
@loadNonDeterministicFiles(
 "C:\calculator\main\--data\nonDeter_to_DeterMethod.txt")
program

    package treplica at "C:\projeto\lib\treplica"
    package cyan.math at "C:\Cyan\lib\cyan\math"
    package main
    package other
\end{lstlisting}
Package \srcstyle{treplica} imported in the first line has metaobject 
\srcstyle{loadNonDeterministicFiles} 
 used in line 2. This metaobject reads the rule files given as parameters (any number of them) and creates a data structure with the loaded data. This data structure is retrieved by metaobject \srcstyle{treplicaAction} in methods \srcstyle{semAn\_codeToAdd} and \srcstyle{afterSemAn\_checkDeclaration}. The first is responsible to replace a nondeterministic method by a deterministic one. The second checks if no other metaobject added a nondeterminitic call in phase semAn (this code addition would not be visible to method \srcstyle{semAn\_codeToAdd}).

For each nondeterministic  method, a rule file defines
a  replacement  method.   This  method   can  be  implemented  by  the
application  of  by   a  supporting  library  and   should  provide  a
deterministic  version of  the indicated  method. For  example, it  is
possible to define  a method that returns not the  current time, but a
timestamp prerecorded in the action object.

\begin{figure}[htbp]
\centering
\begin{lstlisting}[language=Java]
packageA,PrototypeA,methodA - packageB,PrototypeB,methodB
\end{lstlisting}
\caption{Example of nondeterminism rule}
\label{fig:ValidateRuleExample}
\end{figure}

In a  rule file, each rule  is defined in  a line and its  format is
shown in Figure~\ref{fig:ValidateRuleExample}.  The rules are split into
two  parts by  the symbol  \srcstyle{-}.  The  first part  defines the
package, prototype, and name of the
nondeterministic method. The method name of a \textit{keyword method} is composed of each keyword followed by its number of parameters. Then, method
\begin{lstlisting}[language=Java]
    func add: Int n, String s  to: Int n
\end{lstlisting}
has name ``\srcstyle{add:2 to:1}''.

The  second part of the line, after \srcstyle{-}, defines  the
package, prototype, and name of the
deterministic method that will  replace the nondeterministic one.

During  compilation,  metaobject \srcstyle{treplicaAction}
will  replace \srcstyle{methodA}  with a  call to  \srcstyle{methodB},
with the  parameters of the original  call used as parameters  of this
new  call. The object that receives message \srcstyle{methodB} is 
\srcstyle{packageB.PrototypeB}

For example, suppose the line below is in one of the rule files used by the program.
\begin{lstlisting}[language=Java]
util, MyArray, add:2 to:1 - other, DetArray, always:2 theSame:1
\end{lstlisting}
And a method annotated with \srcstyle{treplicaAction} has the code:
\begin{lstlisting}[language=Java]
  // package util was imported
var MyArray array = MyArray new;
array add: 0, ("aazero" substring: 2) to: (Fat fat: 5);
\end{lstlisting}
Then  \srcstyle{treplicaAction} will replace the message passing of the last line by
\begin{lstlisting}[language=Java]
other.DetArray always: 0, ("aazero" substring: 2)
               theSame: (Fat fat: 5);
\end{lstlisting}
If an expression that is an argument to this message passing has a nondeterministic call, it would be replaced according to the rules. That is, the replacement works as expected.

Annotation \srcstyle{nonDeterministic} can be attached to a method to mark it as nondeterministic. The associated metaobject does some checks, described next.
\begin{enumerate}
\item It checks if all superprototype methods with the same name are also attached to the same annotation.
\item Whenever the method is overridden in a subprototype, the metaobject checks if the overridden method is also attached to annotation\\
    \verb|    | \srcstyle{nonDeterministic}\\
    \noindent Then, by this item and the previous one, all methods with the same name in a hierarchy are either all marked as nondeterministic or none is.
\item Whenever there is a message passing in which the attached method may be called, the metaobject checks whether the current method, the method in which the message passing is, is annotated with ``\srcstyle{@nonDeterministic}'' or ``\srcstyle{@treplicaAction}''. If not, it issues an error because the method is also nondeterministic, not marked as such, and the message passing will not be replaced by a deterministic method call.
\end{enumerate}
Metaobject \srcstyle{treplicaAction} uses the data from the rule files to replace some message passings by other ones. It does so even if the method to be replaced is not annotated with \srcstyle{nonDeterministic}. However, if a method is annotated with \srcstyle{nonDeterministic} and no replacement for it is specified in any rule file,  \srcstyle{treplicaAction} issues an error.

There are limitations in the detection of nondeterminism described in this Subsection. It depends on humans to annotate methods as nondeterministic and to put the replacement method call in rule files. There is no help from tools other than metaobject \srcstyle{nonDeterministic}. A tool that does static analysis of the code could detect nondeterminism based on any direct or indirect use of some methods for input and output, time, network, etc. This will be future work.

Rule files may be inconsistent and this is not detected by the metaobjects. The inconsistent arises when the rule files describe the replacement of some methods of a hierarchy but not for others with the same name. Using the previous \srcstyle{MyArray} example, suppose we have the message passing:
\begin{lstlisting}[language=Java]
var MySuperArray array = MyArray new;
array add: 0, ("aazero" substring: 2) to: (Fat fat: 5);
\end{lstlisting}
Assume that \srcstyle{MySuperArray} is superprototype of \srcstyle{MyArray} and that there is no replacement, in the rule files, for method ``\srcstyle{add:2 to:1}'' of \srcstyle{util.MySuperArray}. Then the message send of the last line will not be replaced, even though it is nondeterministic according to one of the lines of a rule file. Metaobject \srcstyle{treplicaAction} is not smart enough to detect this nondeterminism.

A complete solution to this kind of problem demands the knowledge of AST of the whole program. Then a metaobject would do a static analysis in the whole program and detect all nondeterminism based on the \textit{root} nondeterministic methods pointed out in the rule files or annotated with \srcstyle{nonDeterministic}. For example, a method not cited in any rule file and not annotated with \srcstyle{nonDeterministic} would be considered nondeterministic if it may call a nondeterministic method.\footnote{Note that the definition of ``nondeterministic method'' is recursive: a method is nondeterministic if a) it is cited in a rule file or annotated with \srcstyle{nonDeterministic} or b) it may call a nondeterministic method.} Currently, that cannot be done because the MOP limits the information given to metaobjects. These limitations prevent some bad Software Engineering practices but also the implementation of some useful tools. See Guimarães \cite{thecyanmopthesis:2019} for more details.

\section{Case Study: Warriors and Mages}
\label{casestudy}

Active replication allows  us to create applications  that share state
among  different   instances  without   a  central   data  repository.
Peer-to-peer games, in particular  turn-based ``board'' games, are the
perfect  match for  active replication.   These games  have a  complex
state that need to be shared,  composed of the game board, the pieces,
the position  of each  piece, the current  player, among  others. This
state  changes  after each  player's  turn  as  a consequence  of  her
actions, and must be propagated to  the other players before they take
their turns. We created a turn-based  game using Cyan and Treplica, to
assess the  suitability of  creating a replicated  application through
the set of metaobjects presented in this paper.

The two-player board game \emph{Warriors and Mages} simulates a battle
for dominance  of the board.   Each player  controls a group  of three
characters (two warriors and a mage), and players take turns moving or
attacking.   In her  turn a  player can  make three  of these  actions
before the other  player's turn begins. The game ends  when one player
eliminates all the  characters of the opposing  player.  As validation
of the proposed metaobjects and as  an example of how a designer would
use them, we have implemented Warriors and Mages using Cyan and a very
simple  text-based user  interface. In  this section  we will  briefly
describe the  software architecture  of our  implementation as  it was
first designed:  a centralized, nonpersistent application.   Then, we
will show  the simple  steps required  to turn  it into  a replicated,
fault-tolerant application.  The source  code of the final application
can          be         found          in         the          project
tree\footnote{\url{https://bitbucket.org/gdvieira/cyan.treplica.git}}
under \srcstyle{app/rogue}.

\subsection{Implementation Outline}

The  implementation  of the  board  game  follows  a very  simple  MVC
software  architecture.  The  model  is implemented  by the  prototype
\srcstyle{Board}. The  player commands  (move or  attack) are  sent as
method calls  to \srcstyle{Board},  that executes them  accordingly to
the rules of the game.  For instance,  a player is not allowed to make
an   illegal  move.   The   execution  of   the   move  by   prototype
\srcstyle{Board} updates the internal state  of the game, allowing the
correct execution of further moves.

\begin{figure}[htbp]
    \centering
    \includegraphics[width=11cm]{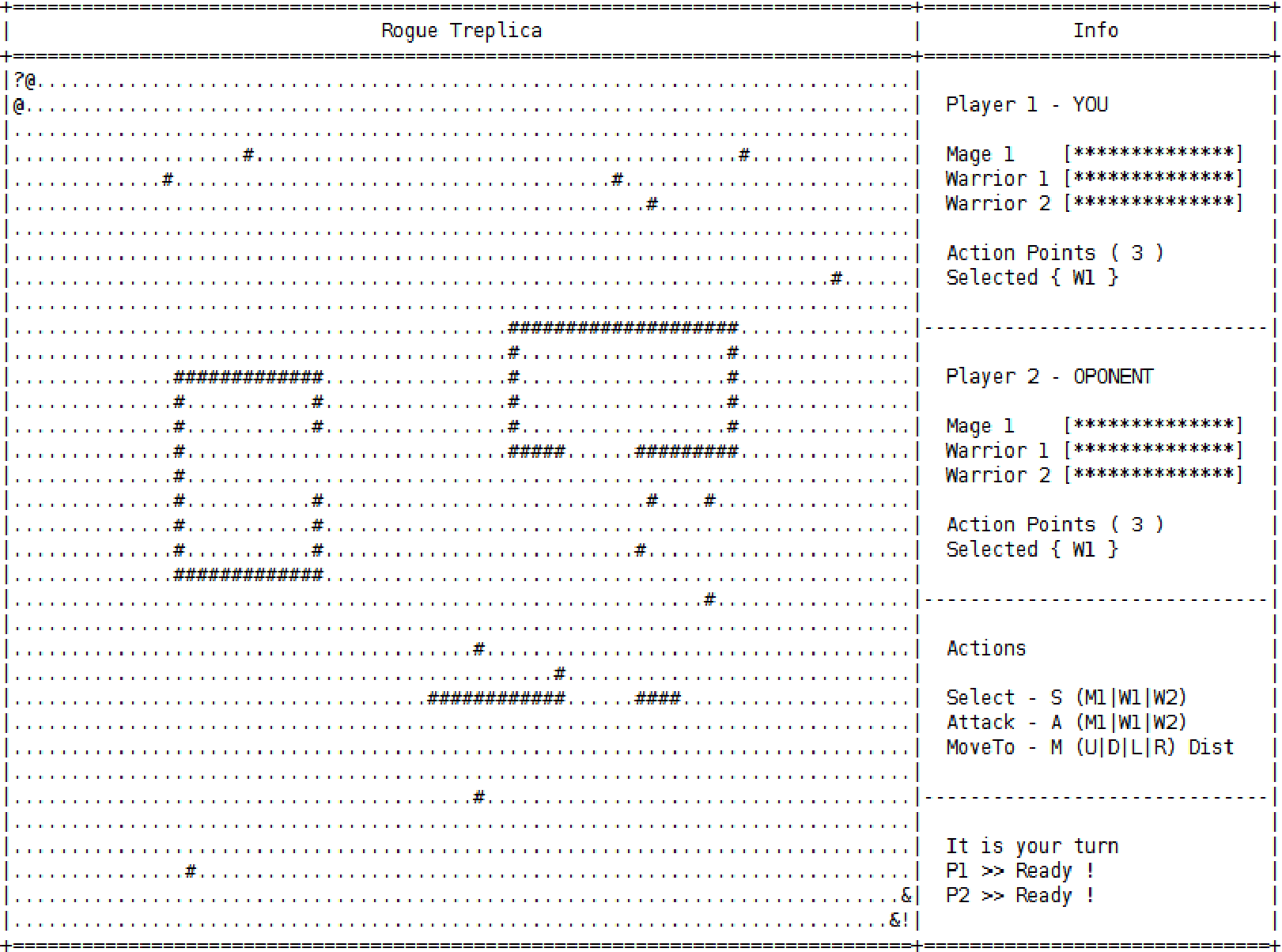}
    \caption{The game board}
    \label{fig:board}
\end{figure}

The prototype  \srcstyle{Window} implements  the view, using  a simple
Java Swing text field containing  a visual representation of the board
as shown in Figure~\ref{fig:board}.   Each letter or symbol represents
an  element  of  the  board.    A  (\srcstyle{.})  is  free  space,  a
(\srcstyle{?}) is a  mage, a (\srcstyle{@}) the warrior  and walls are
represented by a (\srcstyle{\#}). Besides showing the current state of
the board,  the prototype  \srcstyle{Window} receives  keyboard inputs
and passes  them to the  controller.  The commands are  simple letters
(S, M, A), converted to select, move or attack, respectively.

The controller  is implemented  by prototype  \srcstyle{Program}. This
prototype creates the board and the window, acting as a bridge between
the two. It is responsible  for the initialization of the application.
Also, it  receives the callbacks from  prototype \srcstyle{Window} and
calls appropriate methods of prototype \srcstyle{Board}.

\begin{figure}[htbp]
    \centering
    \includegraphics[width=11cm]{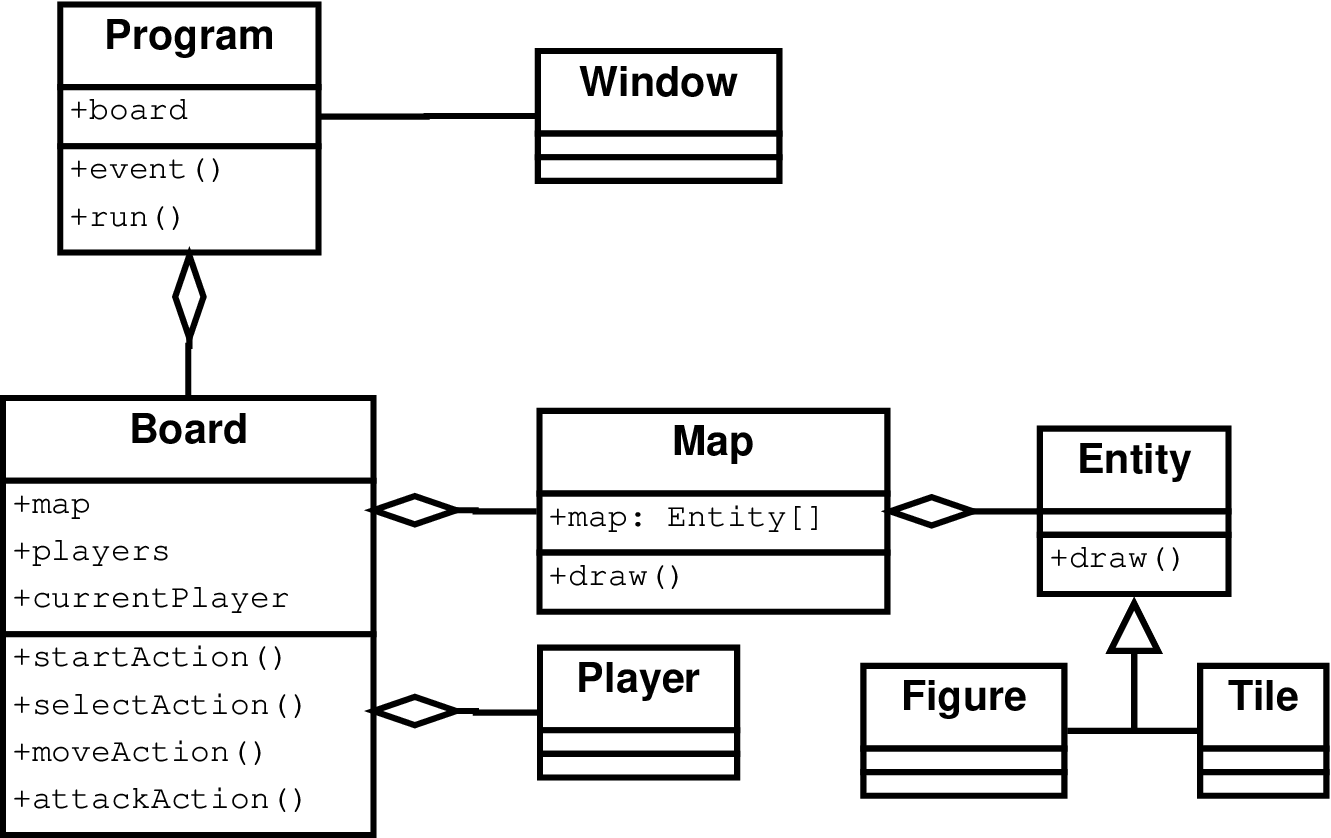}
    \caption{Game implementation main prototypes}
    \label{fig:prototypes}
\end{figure}

To implement the  complete game 14 prototypes  were created, including
these three  central prototypes  (\srcstyle{Board}, \srcstyle{Program}
and    \srcstyle{Window}).    Figure~\ref{fig:prototypes}    shows   a
simplified UML diagram of the  main application's prototypes.  To give
a  sense of  the  style and  complexity of  the  application, we  will
briefly describe the most important prototypes and their relation.

Prototype  \srcstyle{Board} holds  the game  state and  implements its
rules. The  rules are implemented  by key methods  of \srcstyle{Board}
that check if the  a player move is valid and change  the state of the
board accordingly.  These central methods are \srcstyle{startAction:},
\srcstyle{selectAction:},          \srcstyle{moveAction:}          and
\srcstyle{attackAction:},  which  allows  to  reset the  game  to  its
initial  configuration, select  a piece,  move a  piece or  perform an
attack,  respectively.  Other  accessory methods  of \srcstyle{Board},
such    as    \srcstyle{getPlayer:},   \srcstyle{checkEndTurn:}    and
\srcstyle{checkValidTarget:},  allow  checking game  rules  conditions
without changing the local state.

The current state of the game  is held by the \srcstyle{map} attribute
of \srcstyle{Board},  an instance  of prototype  \srcstyle{Map}.  This
prototype holds a matrix with the positions of all elements present on
the board,  each of  them a  sub-prototype of  \srcstyle{Entity}.  The
prototype \srcstyle{Entity}  abstracts the  type of each  element and
also  houses  the  abstract \srcstyle{draw}  method,  responsible  for
creating a  representation of  each element on  screen. There  are two
concrete  sub-prototypes of  \srcstyle{Entity}: \srcstyle{Figure}  and
\srcstyle{Tile}. The prototype \srcstyle{Figure} represents a piece on
the  board,  either  a  mage  of  a  warrior.   This  prototype  holds
information  about   the  current  status   of  each  piece   and  its
capabilities,  providing  methods  to  query  and  change  this  state
respecting the  games rules. The prototype  \srcstyle{Tile} represents
the floor and obstacles that define the playable board.

Finally,   the   prototype   \srcstyle{Board}  keeps   an   array   of
\srcstyle{Player} instances and an attribute \srcstyle{currentPlayer},
allowing it to keep  track of which player is allowed  to play at each
turn.

\subsection{Replication Using Metaobjects}

Once the  application was built,  we wanted  to enable two  players in
different  machines using  replication.  We  assumed the  simple model
where each player executes a local  instance of the game, which stores
a replica of  the game state.  Players take turns,  and the moves each
player executes give the flow of  the game: after three moves are made
by a player, it is the  other player's turn.  Each player executes her
moves as operations that change the local state of the game, and these
operations  should  be  ordered  and replicated  among  all  replicas,
ensuring the same game state for all players.

To  replicate the  game we  are only  interested in  the model  of the
application.  View  and controller  are local to  a replica  and their
state  does   not  need  to   be  replicated.   Thus,   the  prototype
\srcstyle{Board}  is  the  main  locus of  replication,  defining  the
Treplica  context.   This  prototype   was  made  a  sub-prototype  of
\srcstyle{Context} to signal this fact. Once the context is defined we
must identify  the mutator  methods that change  the state  defined by
this context. In  this application these are the  methods of prototype
\srcstyle{Board} that implement player moves (\srcstyle{startAction:},
\srcstyle{selectAction:},          \srcstyle{moveAction:}          and
\srcstyle{attackAction:}).   These  methods  must   be  wrapped  in  a
prototype   that  extends   prototype  \srcstyle{Action}   defined  by
Treplica.   In  our  implementation  that  uses  metaobjects  this  is
accomplished by  the use of the  annotation \srcstyle{@treplicaAction}
in each of these methods, as shown in Figure~\ref{fig:boardproto}.

\begin{figure}[htbp]
\centering
\begin{lstlisting}[language=Java]
object Board extends Context
  ...
  @treplicaAction
  func selectAction: String target { ... }

  @treplicaAction
  func attackAction: String target { ... }

  @treplicaAction
  func moveAction: String direction, String value { ... }
  ...
  @treplicaAction
  func startAction { ... }
  ...
end
\end{lstlisting}
\caption{Prototype \srcstyle{Board}}
\label{fig:boardproto}
\end{figure}

Once the  context and  actions are properly  defined, it is necessary to
instantiate and  start the \srcstyle{Treplica} object  responsible for
replication, binding it to  the context.  Prototype \srcstyle{Program}
is  responsible  for  initialization   of  the  application  including
instantiation of the \srcstyle{Board} object acting as context. At the
point  of  creation of  \srcstyle{Board}  the  use of  the  annotation
\srcstyle{@treplicaInit}  does  all  the required  initialization  and
binding, as shown in Figure~\ref{fig:programproto}.

\begin{figure}[htbp]
\centering
\begin{lstlisting}[language=Java]
object Program extends Input
  ...
  func run: Array<String> args {
    Window build: self;

    var local = ("/var/tmp/magic" ++ args[0]);
    @treplicaInit( 2, 200, local )
    var data = Board new: args[0];
    data startBoard;
    ...
  }
end
\end{lstlisting}
\caption{Prototype \srcstyle{Program}}
\label{fig:programproto}
\end{figure}

Once the  binding of Treplica  and context is  done, all calls  to the
mutator   methods   of    prototype   \srcstyle{Board}   tagged   with
\srcstyle{@treplicaAction} will be replicated.  This fits the original
behavior  of  the  application,  in  which  callbacks  from  prototype
\srcstyle{Window} are  received by \srcstyle{Program} and  turned into
calls  to methods  of \srcstyle{Board}.   Thus, the  main loop  of the
application  remains the  same, moves  are  made by  the local  player
through events generated by the text window. Moreover, board state can
also change though replicated method  calls originating from the other
player's replica.   These moves are  represented in the  local replica
window  as they  happen,  through the  calls  to the  \srcstyle{draw:}
methods of \srcstyle{Map} and \srcstyle{Entity}.

This example shows  that the proposed set of metaobjects  is enough to
allow the replication of a fairly complex application. Moreover, it is
a   demonstration  of   the  simplicity   of  use   of  the   proposed
metaobjects. For a MVC application,  which already has a clear defined
model with a set  of mutator methods, it is only  necessary to add the
proper  annotations.  Other  type  of applications  may  require  some
refactoring effort to define a clear  bounded model. It is part of our
ongoing research  the evaluation of  the ease  of use of  the proposed
metaobjects for general software  projects.  This future work includes
avoidance of some pitfalls, such as nondeterminism.

\section{Related Work}
\label{related}

OpenReplica~\cite{altinbuken2012commodifying}   is   a  framework   to
implement  replicated services  similar to  Treplica~\cite{vieira08a}.
Along with
Treplica,  OpenReplica represents  the  state of  the  art for  easily
creating   replicated   applications   and    both   use   a   similar
object-oriented approach  that suffers from the  same transparency and
code verification problems. Both frameworks require an interface layer
to  encapsulate the  methods  implementing changes  to the  replicated
state   and  neither   allows   code  inspections   that  search   for
inconsistencies in the implementation of the interface.  In this paper
we use metaprogramming  to tackle these challenges,  similarly to the
way metaprogramming has been used to attack similar problems.

Rentschler  et al.~\cite{rentschler2014designing}  argues  the use  of
domain specific  languages (DSLs) to increase  programmer productivity
and  quality and  proposes the  use of  metaprogramming to  translate
these    DSLs   in    other    languages.    They    use   the    Xtend
language~\cite{rentschler2014designing}  to  transform   a  DSL  using
active  annotations.  We  use  a similar  approach  of automatic  code
transformation. However, starting from centralized code written in a general
purpose language,  we arrive  in distributed code  written in  the same
language.
Another  similar work is the one by Blewitt et al.~\cite{blewitt2005automatic} that
proposes  the   use  of   metaprogramming  to   automatically  create
components that implement design patterns.

Xtend~\cite{rentschler2014designing},     Groovy    \cite{groovy2017},
Nemerle   \cite{kamil2005}, Scala \cite{Scala:compiler:plugin}, Rust \cite{Carol:2018:Online}, Python \cite{ramalho2015fluent}, Java \cite{javaplugin:2019:Online}, AspectJ \cite{DBLP:conf/ecoop/KiczalesHHKPG01},
Converge \cite{Tratt:2005:CMD:1146841.1146846}, Elixir \cite{elixir:2018:Online}
and  Cyan~\cite{guimaraes2013cyan}   are
examples of languages with compile-time metaprogramming features.  Some of these languages (Xtend, Groovy, Nemerle, Scala, Java)
allow the traversing  the abstract syntax tree (AST) both to gather
information and to change it. Others use a refined form of Lisp-like \cite{Seibel:2012:PCL:2339396} quotes and unquotes, as Xtend, Nemerle, Groovy, Converge, and Elixir. Languages that only support quotes and unquotes are not directly related to the Cyan MOP.

Compilers of Scala, Java, and Rust support \textit{compiler plugins} that play the same role as metaobjects in Cyan. The documentation about them is scarce except for Scala. Compiler plugins usually can register themselves with the compiler to be called at specific compilation phases. In Cyan, this registration is made by implementing interfaces associated with the phases. For example, a method \srcstyle{semAn\_codeToAdd} is called in phase semAn because the metaobject class implemented interface \srcstyle{IAction\_semAn}. In a compiler plugin, there may not be a direct association between annotations and the plugin. The plugin may take actions when it finds an annotation but, if the compiler plugin is not used, there is no action associated with it (and no compiler error).

AST transformations are similar to the Cyan metaobjects with one difference: they cannot be called when an operation, as subprototyping, method override, or field access is made. In Cyan, some metaobject methods intercept these operations.

The main difference between Cyan and languages with compiler plugins (Scala, Java, Rust) and AST transformations (Xtend, Groovy, Nemerle, Rust, Java) is the way new code is added to a prototype or class. In Cyan, the new code is returned, as a string, by metaobject methods. For example, method \srcstyle{afterResTypes\_codeToAdd} of a metaobject can return a string with fields and methods to be added to the current prototype. The compiler is responsible for adding the new code and it tags the code with the annotation that created it. If there is an error, the compiler will point out exactly who made it. As an example, suppose two metaobjects ask for the insertion of field ``\srcstyle{address}'' to a prototype. Since there will be two fields with the same name, the compiler will issue an error. It will point exactly the two annotations that produced the code.
In all other languages cited above, the code to be inserted should be an AST object.


The language may have macros or language mechanisms that transform strings into AST objects, a step that is not necessary in Cyan. In all other languages, code is added by AST handling. ASTs are a low-level structure of the compiler. They are subject to change, which would invalidate the compiler plugins or AST transformations, and they are not easy to understand. A typical compiler may have more than one hundred AST classes, each one with dozens of methods that can only be understood after a substantial understanding of the compiler. Cyan also allows the visiting of AST nodes, but that is made with the MOP AST, which is a read-only wrapped version of the compiler AST. This AST reflects the language structure, it does not have any specific details of the compiler. And there are security checks when accessing AST nodes. To cite one of them, a metaobject of an annotation in a prototype \srcstyle{P} cannot access the AST of statements of a method in another prototype \srcstyle{T} . This would make prototype \srcstyle{P}  dependent on \srcstyle{T}, a dependency that is not registered by the compiler. If \srcstyle{T} changes, prototype \srcstyle{P}  should be re-compiled. This breaks modularity because internal changes in a prototype are causing the recompilation of another prototype.

It is easy to make a mistake and invalidate the AST when handling it directly. In this case, the compiler may crash, generate wrong output code, or it may detect the invalid AST in a later compilation phase producing a confusing error message. If the code inserted causes a compilation error, the compiler will point out the error. But it will not give any information on which AST transformation or compiler plugin inserted the wrong code.
In general, it is easy to code metaobjects in the Cyan MOP because they do not interact with the original compiler AST and they produce code as strings.

The metaobject \srcstyle{treplicaAction} could be implemented in languages that support compiler plugins and AST transformation. However, there is a feature of this metaobject difficult to implement:
the checks made by metaobject \srcstyle{treplicaAction}, in phase afterSemAn, in the Cyan method annotated with \srcstyle{treplicaAction}.
The metaobject looks for nondeterministic method calls that may have been added by other metaobjects. In this phase, the method code cannot be changed by any other metaobject.

In all other languages, in general there is no guarantee that compiler plugins or AST transformations will not change the code after a compilation phase. Then the checks can be made but, after it, another compiler plugin or AST transformation can add, in the annotated Cyan method, a call to a nondeterministic method.
It may be possible to implement nondeterminism checks using some clever compiler trick, but we are not aware of that. A complete comparison between Cyan and other languages that support metaprogramming is made by Guimarães \cite{thecyanmopthesis:2019}.

Regarding  the problem  of separating  the nonfunctional  requirements
from the functional ones, there are works on AOP (aspect-orientated
programming) that
address the same  problem. AspectJ \cite{kiczales01overview} \cite{aspectj:2018:Online} is a
Java extension  that supports \textit{aspects}. An \textit{aspect} is composed of \textit{pointcuts} and \textit{advice declarations}.
\textit{Pointcuts} pick out points of code called \textit{join points} which may be method calls, object creation, field access, etc. An \textit{advice} is composed of a pointcut and code. It uses the pointcut to select some \textit{join points} that  are wrapped with the advice code.
For example, a \textit{pointcut} can pick out all execution join points related to methods of a Java class  called \texttt{Animal} that start with \texttt{get}.
Then, the \textit{advice} can wrap each method call with code to be executed before, after, or around the call. In AspectJ, this wrapping, called weaving, can happen at compile-time, post-compile time (using the binary files), or at loading time.
With \textit{Intertype declarations}, aspects can add fields and methods to classes and interfaces.

AOP and the Cyan MOP have largely different goals and this fact reflects in their features.
In Cyan, metaobjects can only change  the prototype they are associated with or, in a limited way, calls to methods of the prototype they are associated with.
Aspects can  change multiple files spread in the program, breaking modularity. It is not enough to read a source file in order to understand it because aspects, declared in other files, can change it. This 
is one of eight problems with metaprogramming that the Cyan MOP addresses total or partially \cite{thecyanmopthesis:2019}.
AspectJML \cite{10.1145/3125374.3125383} is an AOP language that solves this problem by associating aspects with types. Each aspect only changes the subtypes thus limiting the affected code.

In Cyan, an annotation attached to the program or to a package (remember annotation \srcstyle{loadNonDeterministicFiles}) can add code to all prototypes of the program or to all prototypes of a package. The annotation is repeatedly applied to all prototypes of the program or the package.
Using this mechanism, many features of AspectJ can be simulated in Cyan. A DSL attached to the annotation can even define \textit{pointcuts} and code that should run on them.

Cyan metaobjects can visit the AST nodes of a prototype, method, statement, etc. This is not possible in AspectJ. Then, aspects cannot detect nondeterminism as metaobject \texttt{treplicaAction}. This detection depends on AST traversal. An aspect can add a superclass or add an implemented interface to a class. In Cyan, metaobjects cannot do that. This was on purpose, to prevent the MOP of making deep changes in the code.

Advice code can use variable \texttt{thisJoinPoint} to access, at runtime, information  on join points. It can also be used to issue errors and warnings at compile-time. However, the information made available by \texttt{thisJoinPoint} is much more limited than in Cyan or any other language supporting compiler plugins or AST transformations. In Cyan, all the information the compiler has that is related to the language itself (and not related to internal compiler details) is available to metaobjects. This is necessary to the metaobject associated with annotation \srcstyle{treplicaAction}.
Figure~\ref{fig:InfoChange} shows the prototype \texttt{Info} modified by annotation \srcstyle{treplicaAction} attached to method \srcstyle{setText:} (see Figure~\ref{fig:TreplicaDadosMeta}). The new methods \texttt{setText:} and \texttt{setTextTreplicaAction:} of Figure~\ref{fig:InfoChange} use information that is available in Cyan, at compile-time, but is not available in AspectJ: the parameter name \texttt{text} (line 3), the prototype and method name used to compose the prototype name \texttt{InfosetText} (line 4), and the method name \texttt{setText:} (used for composing the method name \texttt{setTextTreplicaAction} of line 8). In AspectJ, this information is available only when the final program, changed by the aspects, is running. Hence, in AspectJ, the generated code needs to use reflection and \texttt{thisJoinPoint} in order to access information available at compile-time in Cyan.

Chlipala~\cite{chlipala2013bedrock}   shows  a   proposal  for   using
metaprogramming  to perform  source code  validations at  compile time
using  macros.   Inspecting  the   source  code  for  problems  during
compilation  increases  the  application performance,  because  it  is
unhindered             by            run-time             validations.
Mekruksavanich~\cite{mekruksavanich2012analytical}   proposes  similar
validations  in which  metaprogramming is  used to  detect defects  in
object-oriented programs by the use  of software components capable of
describing and identifying such defects.  Both these works tackle different
problems from  the ones  described in  this paper,  but both  show the
benefits of the use of metaprogramming as an aid in the development of
correct programs.

Compiler  directives, have  been successfully  used to  accelerate the
creation of  parallel programs. OpenMP~\cite{dagum1998openmp}  aims to
ease the  conversion of legacy  centralized C++ and Fortran  code into
portable shared-memory parallel code. OpenACC~\cite{wienke2012openacc}
uses the same approach of compiler directive annotated code to offload
some  compute   intensive  tasks   to  accelerator  devices   such  as
general-purpose  graphic processing  units (GPGPUs).   Both approaches
simplify  the task  of producing  parallel code,  but still  require a
considerable knowledge  of the programmer about  how parallel programs
work. We  use metaobjects in a  simpler way and  aim to completely
shield  the programmer  from  details about  the distributed  programming
model that is used.  Currently we block the occurrence of invalid
nondeterministic   method  calls   and  intend   in  the   future  to
extend detection to other types of consistency violations.



\section{Conclusion}
\label{conclusion}

We have  shown how  to use the  metaprogramming infrastructure  of the
Cyan language to transparently  generate and validate integration code
that  uses the  Treplica  replication framework.   This way,  programs
written  in   Cyan  can  easily   be  converted  from   a  centralized
architecture to a  replicated one by attaching  metaobjects to mutator
methods.  The  set of  metaobjects created showed  for the  first time
that  it  possible  to  automatically  create  replicated  code  using
metaprogramming.

We also demonstrated  the power and simplicity of the  MOP of the Cyan
language  to create  useful metaobjects  in  a very  direct way.   The
programmer of a metaobject in Cyan does  not have to deal with the AST
to  generate  code, she  only  needs  to produce  source  code as strings. However, the programmer can use
the AST  if necessary, for example, to visit the AST nodes of a method.

Moreover, we  demonstrated the potential of  using the metaprogramming
infrastructure  of a  modern language  to make  the use  of frameworks
easier  not only  by removing  boilerplate  code, but  also by  making
semantic validations  that require integration with  the compiler.  In
this paper we validate the generated code with respect to the presence
of nondeterminism, by replacing the nondeterministic operations with
equivalent  deterministic   operations.   We  believe   the  technique
presented to detect nondeterminism can  be extended to other types of
violations of  replication integrity,  such as calling  static methods
outside  the  application context.   In  the  future, we  envision  an
application environment  where distributed programming errors,  one of
the main factors limiting the use of this programming paradigm, can be
directly found by the compiler.  Also, the ability to create isolated,
deterministic operations  seems to be  very useful in the  creation of
tests suites.

\section*{Acknowledgments}

This  work  was  supported  by the  S\~ao  Paulo  Research  Foundation
(FAPESP)  under  grant  \#2014/01817-3  and  by  FIT  ---  Instituto  de
Tecnologia.


\bibliographystyle{elsarticle-num}
\bibliography{cyan.metaobjects.treplica}

\end{document}